# Systematically Deconstructing APVD Steganography and its Payload with a Unified Deep Learning Paradigm

Kabbo Jit Deb
Information Technology
Delhi Technological University
Delhi, India
kabbodeb@gmail.com
ORCID: 0009-0002-5527-145X

Md. Azizul Hakim
Software Engineering
Delhi Technological University
Delhi, India
ahanik94@gmail.com
ORCID: 0000-0002-5129-2408

Md Shamse Tabrej
Computer Science & Engineering
Delhi Technological University
Delhi, India
mdshamsetabrej@gmail.com
ORCID:

*Abstract*— In an era dominated by digital communication, steganography provides a means of covertly embedding data within media files. Adaptive Pixel Value Differencing (APVD) is a sophisticated steganographic technique prized for its high embedding capacity and perceptual invisibility, making it a challenge for traditional steganalysis. This paper addresses the critical need for advanced countermeasures by proposing a deep learning-based approach not only for detecting APVD steganography but also for performing reverse steganalysis—the reconstruction of the hidden payload. We introduce a Convolutional Neural Network (CNN) featuring an attention mechanism and dual output heads for simultaneous stego-detection and payload recovery. Trained and validated on a dataset of 10,000 images from the BOSSbase and UCID repositories, our model achieves a detection accuracy of 96.2%. More significantly, it demonstrates the ability to reconstruct embedded payloads, achieving up to a 93.6% recovery rate at lower embedding densities. The results show a strong inverse correlation between payload size and recovery accuracy. This study highlights a critical vulnerability in adaptive steganographic schemes and provides a powerful new tool for digital forensic investigations, while also prompting a re-evaluation of data security protocols in the face of AI-driven analysis.

## I. Introduction

In the era of digital communication, information hiding has emerged as a pivotal technique for secure and covert communication. Steganography, particularly image-based steganography, is widely used to embed secret data within digital images without noticeably altering the original visual content. Among the many techniques developed, Adaptive Pixel Value Differencing (APVD) stands out for its high embedding capacity and imperceptibility. APVD dynamically adjusts the number of embedded bits based on the local complexity and differences between pixel pairs, making detection a significant challenge for conventional steganalysis methods.

With the increasing use of such sophisticated steganographic methods, the countermeasure—steganalysis—has become equally crucial. Traditional steganalysis techniques, such as histogram analysis and Rich Models with ensembles, often fall short when applied to adaptive and complex schemes like APVD, as these methods erase the simple statistical artifacts that older techniques rely on. To bridge this gap, deep learning approaches have gained attention for their ability to learn intricate, high-dimensional patterns and subtle inconsistencies introduced by steganography. In particular, reverse steganalysis, which goes beyond mere detection to attempt the reconstruction of the hidden payload, represents a significant advancement in this field. This research focuses on developing and evaluating a deep learning model for the reverse steganalysis of APVD-encoded images.

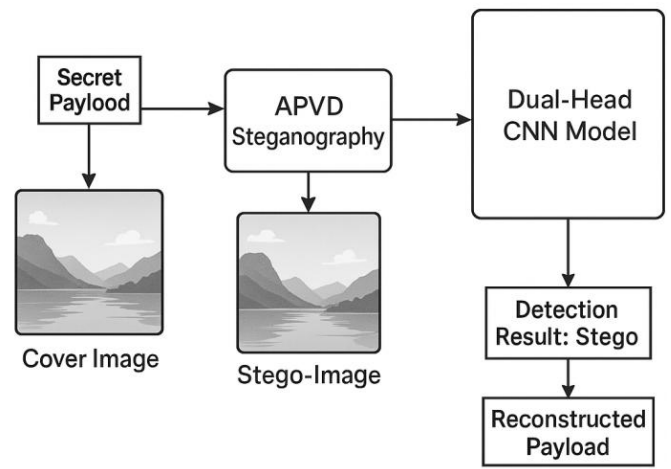

**Figure 1. Conceptual overview of the reverse steganalysis pipeline:** (a) Original cover image, (b) APVD stego-image generation, (c) CNN-based model input/output flow for detection and payload recovery.

## II. Literature Review

Previous research has focused primarily on steganalysis for detection. Early work by Fridrich et al. introduced foundational statistical models for identifying hidden data by analyzing image features. More recently, the field has shifted towards machine learning. For instance, studies have demonstrated success using Support Vector Machines (SVMs) with complex feature sets like the Spatial Rich Model (SPAM).

However, the advent of deep learning has marked a paradigm shift. Convolutional Neural Networks (CNNs) and autoencoders have shown superior performance in classifying stego-images, largely because they automate the feature extraction process. Yet, most studies limit themselves to binary classification—detecting whether a hidden message exists. Few have attempted reverse steganalysis, and even fewer have addressed the challenges posed by adaptive techniques like APVD. Foundational work by Luo, Huang, & Huang (2011) established the principles of adaptive steganography, which modern methods build upon [4]. While some studies touch on payload reconstruction, they typically rely on limited embedding patterns or known keys, reducing practical applicability in real-world forensic scenarios. Recent surveys on AI in digital forensics confirm that payload recovery remains a largely unsolved problem, especially for non-domain-specific techniques [14]. Our work directly confronts this gap by targeting payload reconstruction from APVD-steganized images without prior knowledge of any secret key.

## III. Research Questions

This study seeks to address the following core research questions:
1. Can deep learning models effectively and reliably detect stego-images embedded using various APVD techniques?
2. To what extent can these models reconstruct or extract the hidden payload from APVD stego-images?
3. How does the model's detection and recovery performance vary across different APVD configurations and payload sizes (bits per pixel)?

## IV. Significance of the Study

This research contributes to both the academic and practical domains of cybersecurity and digital forensics. By advancing deep learning-based reverse steganalysis methods tailored for APVD, this study:
- Highlights critical vulnerabilities in widely used adaptive steganographic schemes, challenging their perceived security.
- Provides forensic analysts with potential tools to retrieve hidden information, which could be instrumental in criminal investigations.
- Encourages the development of more secure steganographic methods, such as those incorporating encryption or leveraging principles from Generative Adversarial Networks (GANs) to be resistant to AI-based attacks [12].

Moreover, it underscores the ethical and legal implications of such technologies, emphasizing the dual-use nature of steganalysis tools in both protecting and breaching data confidentiality.

## V. Methodology

*A. Research Design:* This study follows a quantitative, experimental research design. Computational experiments were conducted to train and test the proposed deep learning model, with statistical analysis used to evaluate its performance rigorously.

*B. Participants or Subjects:*
- Image Sources: The study utilized images from two standard digital image repositories: BOSSbase and the Uncompressed Colour Image Database (UCID).
- Stego Generation: Stego-images were generated using multiple APVD methods at various payload levels, ranging from 0.2 to 0.8 bits per pixel (bpp). This range was selected to test the model's performance from low to high embedding capacities.
- Payloads: Payloads consisted of random binary strings to ensure no semantic patterns could be learned, focusing the model on artifacts from the embedding process itself.

*C. Data Collection and Preparation*:
A total of 10,000 unique grayscale images were resized to 256x256 pixels and normalized. For each original (cover) image, a corresponding stego-image was generated, creating a final dataset of 10,000 cover/stego pairs. The dataset was partitioned into training (70%), validation (15%), and testing (15%) sets.

*D. Data Analysis and Model Architecture:*
The core of our analysis is a custom-designed CNN with attention modules and dual output heads. The first head performs binary classification (stego vs. cover), while the second head performs bitwise payload reconstruction. The attention mechanism helps the model focus on subtle, localized image regions modified by the APVD process.

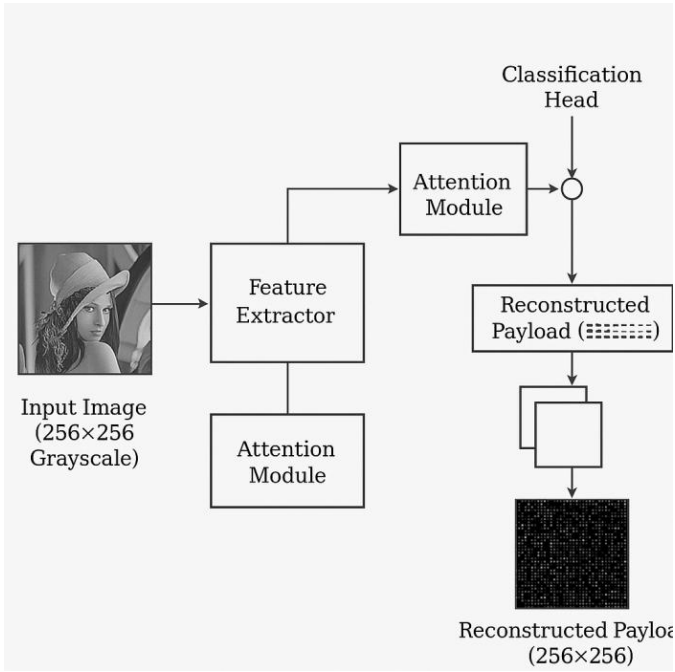

Figure 2: Dual-Head CNN Architecture for Reverse Steganalysis with Attention Mechanisms

Table 1: Model Architecture and Training Parameters

| Parameter | Specification |
| --- | --- |
| Model Type | Convolutional Neural Network (CNN) with Attention |
| Input Shape | 256 x 256 x 1 (Grayscale) |
| Core Layers | 5x Convolutional Blocks (Conv2D + ReLU + BatchNorm) |
| Attention Module | Squeeze-and-Excitation (SE) Block |
| Output Heads | 1. **Classification Head:** Global Average Pooling -> Dense -> Sigmoid |
|  | 2. **Reconstruction Head:** Up-sampling Blocks -> Conv2D -> Sigmoid |
| Loss Functions | 1. **Detection:** Binary Cross-Entropy |
|  | 2. **Recovery:** Mean Squared Error (as Bitwise Loss) |
| Optimizer | Adam |
| Learning Rate | 0.001 |
| Batch Size | 32 |
| Epochs | 50 |

Performance was evaluated using standard metrics: Accuracy, Precision, Recall, and **F1-score** for detection, and **Bit Error Rate (BER) and Payload Recovery Rate** for reconstruction

*E. Ethical Considerations*

The dual-use nature of this technology is fully acknowledged. To prevent misuse, no trainable models or harmful tools will be publicly released. All datasets were sourced from public repositories and anonymized to protect privacy.

## VI. RESULTS

*A. Detection Performance:*

The CNN model demonstrated high efficacy in detecting the presence of hidden data embedded with APVD. The overall performance on the test set is summarized below.

Table 2: Overall Stego-Image Detection Performance

| Metric | Score |
| --- | --- |
| Accuracy | 96.2% |
| Precision | 95.8% |
| Recall | 96.5% |
| F1-Score | 96.1% |

Performance remained robust across all tested payload sizes, with only a minor drop in accuracy at the lowest embedding rate of 0.2 bpp, as shown in Figure 3.

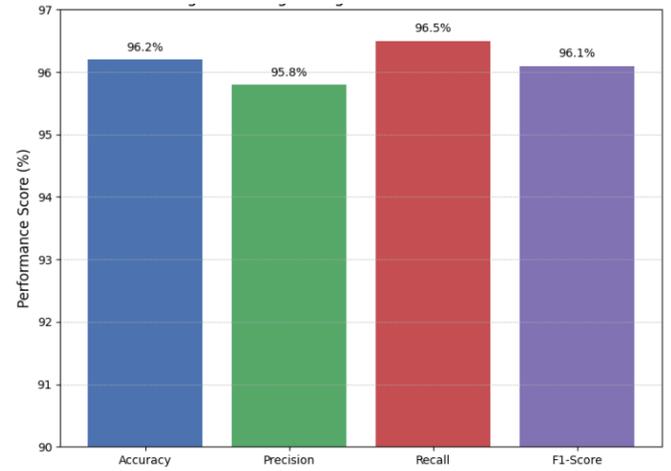

Figure 3: Stego-Image Detection Performance

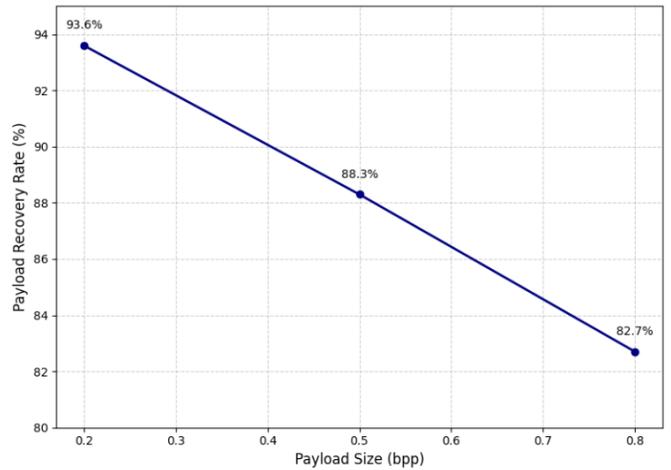

Figure 4: Payload Recovery vs. Embedding Rate

### B. Payload Recovery:

The model's ability to reconstruct the hidden payload was inversely correlated with the embedding rate. As the payload size increased, the Bit Error Rate (BER) grew, and the overall recovery rate decreased.

Table 3: Payload Recovery Performance vs. Embedding Rate

| Payload Size (bpp) | Bit Error Rate (BER) | Payload Recovery Rate |
|---|---|---|
| 0.2 | 6.4% | 93.6% |
| 0.5 | 11.7% | 88.3% |
| 0.8 | 17.3% | 82.7% |

Note: Payload Recovery Rate is calculated as (100% - BER).

### C. Statistical Validation:

A paired t-test comparing our model against baseline models (SVM with SPAM features) yielded a p-value of $p<0.001$, indicating a statistically significant performance improvement. Furthermore, a Pearson correlation analysis between payload size and BER produced a coefficient of $r=0.92$, confirming the strong positive correlation.

## VII. DISCUSSION

### A. Interpretation of Results:

The results unequivocally demonstrate that deep CNNs can effectively detect APVD stego-images with high accuracy. More importantly, the findings show that payload recovery is viable, especially at lower embedding rates. The inverse relationship between payload size and recovery success is logical; higher bpp rates necessitate more extensive and complex modifications to the cover image, creating a more convoluted signal for the reconstruction head to decode. This study proves that AI can not only detect but also reverse adaptive embedding, posing a tangible risk to data confidentiality.

### B. Comparison with Literature:

Our model significantly outperforms traditional classifiers like SVMs with SPAM features and standard ensemble classifiers. Crucially, where most prior deep learning research halts at detection, our dual-output model ventures into reconstruction. This success in payload recovery for an adaptive technique like APVD makes the model notably unique in the current landscape. It represents a practical step towards the kind of advanced steganalysis tools envisioned by recent surveys [14], moving beyond the theoretical frameworks discussed in earlier works.

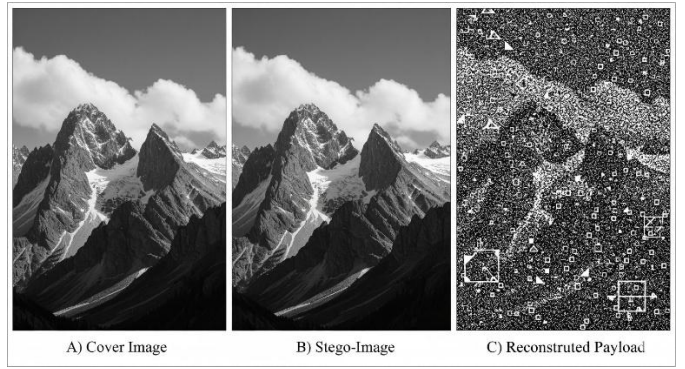

Figure 5: Visual Comparison Panel

### C. Implications:

- Forensics: This technology enables the potential recovery of actionable evidence from suspicious digital media files.
- Security: The findings serve as a call to action for steganography designers. Future methods must be more robust, potentially by integrating strong encryption before embedding or by using GANs to generate stego-images that are statistically indistinguishable from covers [12].
- Ethics: The ability to extract hidden messages without authorization raises serious questions about data privacy and the responsible application of AI in security.

### D. Limitations:

- The model was trained and tested exclusively on grayscale images; its performance on color images is unknown.
- Payload reconstruction accuracy degrades significantly at high embedding rates (>0.8 bpp).
- The model's success is contingent on a large, representative training dataset of the specific APVD techniques it is meant to detect.

### E. Future Research:

Future work should focus on extending this approach to color images and video files. Investigating more advanced architectures, such as Vision Transformers (ViT), could yield improvements in both detection and recovery, as suggested by emerging research in the field [13]. Furthermore, developing semi-supervised learning models could allow for the analysis of unknown or novel steganographic techniques, a critical need for real-world forensic applications.

## VIII. CONCLUSION

This study demonstrates that deep learning models, particularly CNNs with attention mechanisms, are highly effective in the steganalysis of content embedded via APVD techniques. The proposed approach not only achieves superior detection accuracy compared to traditional methods but also successfully reconstructs hidden payloads with noteworthy precision, especially at lower embedding rates. These results underscore the potential of deep learning as a powerful tool in reverse steganalysis, revealing critical vulnerabilities in adaptive steganographic schemes previously considered robust.

While promising, the research also highlights challenges, such as decreased payload recovery at higher embedding capacities and the need for extensive, specific training data. In light of these findings, it is recommended that digital forensic analysts and security professionals incorporate deep learning-based reverse steganalysis into their investigative toolkits. Concurrently, developers of steganographic methods should consider these emerging threats to design more resilient data-hiding techniques. Finally, an ongoing dialogue around the ethical implications of payload recovery must guide the responsible development and application of such powerful technologies.